\documentclass[12pt]{article}
\usepackage{amssymb}
\usepackage{graphicx}
\usepackage{amsmath}

\def\d{\partial}
\def\l{\left(}
\def\r{\right)}

\newcommand{\be}{\begin{equation}}
\newcommand{\ee}{\end{equation}}
\newcommand{\ba}{\begin{align}}
\newcommand{\ea}{\end{align}}
\newcommand{\bg}{\begin{gather}}
\newcommand{\eg}{\end{gather}}
\newcommand{\bseq}{\begin{subequations}}
\newcommand{\eseq}{\end{subequations}}

\def\half{\frac{1}{2}}

\begin{document}

\title{Strong-coupling scale\\
and frame-dependence of the initial conditions \\ for chaotic
  inflation in models \\ with modified (coupling to) gravity}

%\title{$R^2$-inflation with conformal SM Higgs field}
\author{Dmitry Gorbunov, Alexander Panin\\
%\affiliation{Institute for Nuclear Research of Russian Academy of
%  Sciences,\\60-th October Anniversary pr.\;7a, 117312 Moscow, Russia} 
%\affiliation{Moscow Institute of Physics and Technology,\\Institutsky
%  per.\;9, 141700 Dolgoprudny, Russia} 
\mbox{} {\small\em Institute for Nuclear Research of Russian Academy of
  Sciences, 117312 Moscow,
  Russia}\\  
\mbox{} {\small\em Moscow Institute of Physics and Technology, 
141700 Dolgoprudny, Russia}
}
\date{}

\maketitle

\begin{abstract}
A classical evolution in chaotic inflationary models starts at high
energy densities with semi-classical initial conditions presumably
consistent with universal quantum nature of all the fundamental
forces. That is each quantum contributes the same amount to the
energy density. We point out the upper limit on this amount inherent
in this approach, so that all the quanta are inside the weak-coupling
domain. We discuss this issue in realistic models with modified
gravity, $R^2$- and Higgs-inflations, emphasizing  the specific change of
the initial conditions with metric frame, while all the quanta still
contribute equal parts. The analysis can be performed straightforwardly
in any model with modified gravity ($F(R)$-gravity, scalars with
non-minimal couplings to gravity, etc).  
\end{abstract}

%%%%%%%%%%%%%%%%%%%%%%%%%%%%%%%%%%%%%%%%%%%%%%%%%%%%%%%%%%%%%%%%%
%\section{Introduction and summary}

{\it 1.} The idea of early time inflation---an epoch when the Universe expands
almost exponentially---has been put 
forward\,\cite{starobinsky,Guth:1980zm,Linde:1981mu,Albrecht:1982wi} 
to solve the problems of initial conditions in the Hot Big Bang
theory. Namely, the Universe becomes as we know its---flat,
homogeneous, isotropic and populated with adiabatic scalar
perturbations---at the inflation preceding the hot stages. 
However, a particular model of inflation can be recognized as one
truly solving the initial condition problems only if realization of
the exponential expansion occurs naturally or likely: may be a special, but
reasonably large part of the initial phase space of the dynamical
variables leads to the sufficiently fast expansion. 

Evidently, the issue of naturality includes a task of outlining the
available part of the model phase space to start the evolution 
from. The other tasks are estimating the relative probability of
beginning from a 
particular point in the phase space and evaluating the subsequent
dynamics. The higher the probability {\it to start the evolution ending in
inflation,} the more natural the given inflationary model. 

The naturality issue can be properly addressed in models of chaotic
inflation\,\cite{Linde:1983gd,Linde:1984ir}. 
There the dynamics starts from initial conditions attributed to the
semi-classical description of quantum processes in a spatial cell of
the Planck-scale size. The assumption is that all the
ingredients---like curvature for gravity and potential, kinetic, and
gradient energies for scalar fields---give roughly the same amount to the
total energy density. The amount is naturally expected to be of order
one in Planck units. In each cell these quantities fluctuate around
the Planck value. If accidentally the potential term becomes somewhat
bigger than others, the cell expands, the energy density decreases,
but the contributions of kinetic and gradient terms do it much faster,
leaving the region dominated by the scalar 
potential \cite{Albrecht:1985yf,Albrecht:1986pi,East:2015ggf}. 
Then the energy density dominates, which makes the expansion fast, almost
exponential. 

{\it 2.} 
This scheme works perfectly for the power-law potentials like original 
$\lambda\phi^4$\,\cite{Linde:1983gd}, 
but seemingly fails for inflation models with flat, plateau-like 
potential favored by the present cosmological data. 
% Ref on Planck
The value of the scalar flat potential at inflation is constrained
from above by the negative searches for imprints of the relic
gravitational waves---tensor perturbations arisen at the inflationary
stage. The allowed value is much below the Planck scale, hence the
beginning with all the terms being of order one in Planck units is
simply impossible in these models. 
If the gradient and the kinetic terms dominate, the
Universe expands but moderately, insufficiently for initiating an 
inflationary stage and hence for solving the Hot Big Bang problems.  
Only the regions, where the kinetic and gradient terms take 
enormously small values, so that the potential dominates, can evolve
into inflationary regime. However, such a fluctuation is extremely
rare, and hence the chaotic inflation with a flat potential is very
unlikely. 

In particular, one of the most developed models of this
type---$R^2$-inflation\,\cite{starobinsky} 
and Higgs-inflation\,\cite{Bezrukov:2007ep}---were accused of
suffering from this unlikeness\,\cite{Ijjas:2013vea}. 
Recently it has been realized \cite{Gorbunov:2014ewa} 
that, quite the contrary, within the paradigm of chaotic inflation 
what must be treated as very unnatural is the 
situation, when the kinetic and gradient
terms much exceed the potential one. The chaotic inflation paradigm
implies that all the terms must be of the same order, and hence about the
value $V_0$ of the flat scalar potential. Recall, this value is fixed
from the cosmological observations. 

The key point of Ref.\,\cite{Gorbunov:2014ewa} 
is that the hierarchy between the contributions of different
ingredients to energy density remains the same independently of the
metric frame. Since both of the models under discussion exhibit rather
non-trivial coupling to gravity, the recognition of the proper
ingredients (degrees of freedom of the quantum system) is subtle in
the so-called Einstein frame, where the gravity itself is Einstenian
(General Relativity) 
and the scalar potential is flat. The situation becomes much clearer in
the Jordan frames, where the models have been originally
formulated. There, as in any other frames, the properly defined
ingredients give the same amount to the energy density. However, this
amount, related by the Weyl transformation to the fixed by
observations value $V_0$ of the scalar potential, happens to be not
confined to be (right) below the Planck scale. 
Moreover, if in the Einstein frame one takes the gradient, kinetic or
gravity terms to be of the Planck scale, then upon the Weyl
transformation, in the Jordan frame the energy density grossly exceeds
the strong coupling gravity scale.  
This peculiarity
deserves a little investigation, which bring us to the main subject of
this note, that is frame-dependence of the chaotic initial
conditions. 

{\it  3.} 
The starting point here is the motivation for limiting from above 
the term contributions to the total energy density. First, whichever
cell is taken, its spatial scale $\Delta x$ defines the critical 
scale of energy fluctuations $\Delta E$ in it 
through the quantum uncertainty relation $\Delta E
\Delta x\geq 2\pi$. Second, one constrains all the quantities to be
inside the weak coupling domain, otherwise we simply cannot properly
define the quantity itself. For the case of General Relativity and
scalar field (inflaton) minimally coupled to gravity, these two
observations pin down the Planck scale as the critical value. For the
inflationary models with non-minimal (coupling to) gravity one has to
determined the corresponding critical value in a given metric frame
and then check how it changes from frame to frame. 

We perform this study for the Higgs-inflation. We begin with the
Jordan frame (JF), where settling the unitary gauge with Higgs
boson $h$ the model is described by the Lagrangian 
\begin{equation}
\label{JF-Higgs}
S^{\text{JF}}=\int \sqrt{-g^{\text{JF}}} d^4x \left[ 
-\frac{M^2_{\text{Pl}}}{16\,\pi}\l
1+\frac{8\,\pi\,\xi\,h^2}{M^2_{\text{Pl}}}\r R^{\text{JF}} + \half
g_{\mu\nu}^{\text{JF}} \d^\mu h\d^\nu h -\frac{\lambda}{4}h^4
\right]\,.
\end{equation}
The observed matter perturbations fix the parameter of the
non-minimal coupling to gravity at the value about 
$\xi\sim 5\times 10^4$ \cite{Bezrukov:2007ep} 
when the model is considered at the tree-level\footnote{We ignore here
subtleties related to non-renormalizability of the
couplings to gravity, treatment of the higher order corrections 
and behavior of the quantum effective potential
at inflationary scale, 
referring to paper \cite{Bezrukov:2013fka} for a comprehensive review.}.  
A homogeneous (inside a given cell) 
configuration of the scalar field $h$ changes the
effective Planck mass 
\[
M_{\text{Pl}}\to M_{\text{Pl}} \sqrt{1+\frac{8\,\pi\,\xi\,h^2}{M^2_{\text{Pl}}}}
\]
pushing with large value of $h$ the scale of strong gravity up to
higher energies, 
\begin{equation}
\label{2}
E_{\text{GR}}\to \Lambda_{\text{JF}}\equiv \sqrt{8\,\pi\,\xi}h\,.
\end{equation}
Hence, the critical value of energy density, $E_{\text{JF}}^4$, that
is the highest value consistent with the semiclassical treatment of
the whole system, must
not exceed $\Lambda_{JF}^4$. The quantum uncertainty relation then
allows for considering any cell of spatial size $l\geq 
2\pi/E_{\text{JF}}$. 

According to the chaotic inflation paradigm\,\cite{Linde:1983gd} all
the ingredients contribute the same amount to the total energy
density. At large values of $h$ the proportional to $\xi$ term in
eq.\,\eqref{JF-Higgs} defines the gravity, the scalar kinetic and the
scalar gradient contributions to the total energy 
density\,\cite{Gorbunov:2014ewa}. Consequently, the chaotic inflation
implies the following initial conditions
\[
\xi h^2 R^{\text{JF}} \sim \xi \dot h^2 \sim \xi
(\d_i h)^2 \sim h^4\sim E_{\text{JF}}^4 \,.
\]
Thus we arrive at the estimate for the curvature 
$ R^{\text{JF}}\sim h^2/\xi$ and with 
\eqref{2} and $\xi\gg 1$ one confirms that indeed the
inflationary scale $E_{\text{JF}}$ is below $\Lambda_{\text{JF}}$ at
large enough $h$, so
the dynamics develops well inside the weak-coupling domain and the
semi-classical treatment is fully applicable. {\em In the scalar sector}
(the Higgs boson and all the Standard Model fields) {\em the strong
coupling scale is lower than that in the gravity sector, and actually
coincides with $E_{\text{JF}}\sim h$\,\cite{Ferrara:2010yw,Bezrukov:2010jz}, 
which indeed defines the critical
value,} that is minimum of the strong coupling scales in all the
sectors of the model, see Ref.\,\cite{Bezrukov:2011sz} for a brief summary.  

Performing the Weyl transformation, 
\begin{equation}
\label{Weyl}
g_{\mu\nu}^{\text{JF}}\to g_{\mu\nu}^{\text{EF}}= \Omega^2 \,
g_{\mu\nu}^{\text{JF}}\,, \;\;\;\text{with}\;\;\; 
\Omega^2\equiv 1+ \frac{\Lambda^2_{\text{JF}}}{M^2_{\text{Pl}}}\,, 
\end{equation}
we come to the Einstein frame, where pure gravity term is Einsteinian
with gravity scale $M_{\text{Pl}}$. 
It is shown in Ref.\,\cite{Gorbunov:2014ewa} that in this frame all
the terms give equal contributions to the total energy density of 
order 
\[
E_{\text{EF}}\sim \Omega^{-1}\cdot E_{\text{JF}} \sim
E_{\text{JF}}\cdot \frac{M_{\text{Pl}}}{\Lambda_{\text{JF}}}\sim  
\frac{M_{\text{Pl}}}{\sqrt{\xi}}\;.
\]
{\em Note, that the estimate of the critical value $E_{\text{EF}}$ does not
depend on the value of $E_{\text{JF}}$,} while the dynamics is well
inside the weak coupling domain. Recall again, the value of
$E_{\text{EF}}$, that is parameter $\xi$, is fixed from the cosmological
observations. The energy density of the plateau-like scalar potential
at inflation is $V_0\sim
M_{\text{Pl}}^4/\xi^2$\,\cite{Bezrukov:2007ep}. 
The minimum of the strong coupling scales in the Einstein frame coincides
at inflationary stage with $E_{\text{EF}}\sim
M_{\text{Pl}}/\sqrt{\xi}$ \,\cite{Ferrara:2010yw,Bezrukov:2010jz}, so
it is indeed the critical value. The quantum uncertainty relation
$\Delta E\Delta x\geq 2\pi$ implies that in the Einstein frame only a
patch of the spatial size exceeding 
\[
1/E_{\text{EF}}\sim
\sqrt{\xi}/M_{\text{Pl}}\gg 1/M_{\text{Pl}}
\]
 can be treated
within semi-classical approach, providing the initial conditions for 
the classical equations describing evolution of the 
inflaton field and the Universe expansion.  One concludes, that the initial
conditions in the both Jordan and Einstein frames are truly in accord with the
chaotic inflation paradigm.   

{\it 4.} 
Similarly one can address the issue of initial conditions in
$R^2$-inflation\,\cite{starobinsky}. The model is described in the
Jordan frame by the following Lagrangian 
\begin{equation}
\label{R2-JF}
S^{\text{JF}}=-\frac{M^2_{\text{Pl}}}{16\,\pi} 
\int\!\! \sqrt{-g^{\text{JF}}}\, d^4x\; R^{\text{JF}} \l
1-\frac{R^{\text{JF}}}{6\,\mu^2}\r\,. 
\end{equation}
The value of mass parameter $\mu$ is fixed from cosmological
observations as $\mu\simeq 2.5\times 10^{-6}\times M_{\text{Pl}}$. 
Similarly to the previous model an average curvature in a  
given cell changes the effective (local in a sense) Planck mass. The scale of 
strong gravity regime $E_{\text{GR}}$ inside the cell also depends on 
the curvature $R^{\text{JF}}$. For the large enough values of the latter we have 
\begin{equation}
\label{LambdaR2}
E_{\text{GR}}\to \Lambda_{\text{JF}}\equiv \frac{M_{\text{Pl}}}{\sqrt{6}\,\mu}
\sqrt{R^{\text{JF}}}\,.
\end{equation}
For sufficiently large $R^{\text{JF}}$ the first term in the brackets  
in Eq.~\eqref{R2-JF} is negligible. Consequently, the contribution of
the curvature 
generated by initial fluctuation of energy $E_{\text{JF}}$ is
estimated as 
\[
E_{\text{JF}}^4 \sim M_{\text{Pl}}^2\frac{(R^{\text{JF}})^2}{\mu^2}\,,
\] 
which implies the energy scale well below the 
strong gravity scale~\eqref{LambdaR2}.

Now let us look at the theory in the Einstein frame, to which one can come 
by the same  Weyl transformation~\eqref{Weyl} with $\Lambda_{\text{JF}}$ given 
by Eq.~\eqref{LambdaR2}. At this frame the energy of the initial fluctuation is 
\[
E_{\text{EF}}\sim \Omega^{-1}\cdot E_{\text{JF}} \sim
E_{\text{JF}}\cdot \frac{M_{\text{Pl}}}{\Lambda_{\text{JF}}}\sim  
\sqrt{\mu\, M_{\text{Pl}}}\;.
\]
{\em As in the Higgs-inflation model it does not depend on the Jordan frame 
value while its energy density coincides with the value of 
plateau-like inflaton potential during inflation.}  Likewise, 
the quantum uncertainty relation
$\Delta E\Delta x\geq 2\pi$ implies that in the Einstein frame only a
patch of the spatial size exceeding $l\sim 1/\sqrt{\mu\,
  M_{\text{Pl}}}\gg 1/ M_{\text{Pl}}$ can be treated
semi-classically.

%%%%%%%%%%%%%%%%%%%%%%%%%%%%%%%%%%%%%%%%%%%%%%%%%%%%%%%%%%%%%%%%%
%\section{Discussion}

{\it 5.} To summarize, we have analyzed the frame-dependence of the
initial conditions in inflationary models with modified gravity,
emphasizing the importance of being in the weak coupling regime at the
onset of expansion. With examples of the Higgs-inflation and
$R^2$-inflation we illustrated that that the Weyl transformation
between the frames properly changes the conditions (initial
energy density, which is the same for all the terms), 
but keeps the model safely within the weak coupling
domain. Naturally, the initial energy density, being frame-dependent,
must not coincide with the Planck scale $M_{\text{Pl}}$, but rather be
limited by the corresponding strong coupling scale.  This conclusion
seems to be general for the $F(R)$ modified gravity models and models
with scalars non-minimally coupled to gravity, though an
analysis similar to what is done above may be complicated by presence 
of several dynamical scalar fields.

%We can start from any value below the strong coupling scale, but the
%largest vaue is most natural ??
%Tunneling instead of chaotic for the flat potentials? 

\vskip 0.3cm 

We thank Yu.\,Shtanov for raising the issue of why the critical energy
is not always at the Planck scale. The work is supported by the RSF
grant 14-12-01430.

%%%%%%%%%%%%%%%%%%%%%%%%%%%%%%%%%%%%%%%%%%%%%%%%%%%%%%%%%%%%%%%%
%%%%%%%%%%%%%%%%%%%%%%%%%%%%%%%%%%%%%%%%%%%%%%%%%%%%%%%%%%%%%%%%

\end{document}